\documentclass[aps,prl,superscriptaddress
                      ,twocolumn
                                ]{revtex4}

\usepackage{graphicx
	    ,amssymb
	    ,color
	    ,siunitx
	    ,multirow
	    ,booktabs
	    ,amsmath
	    ,array
	    }

\newcommand{\cm}{\,cm$^{-1}$}

\newcommand{\mos}{MoS$_2$}

\newcolumntype{C}{>{$}c<{$}}
\AtBeginDocument{
\heavyrulewidth=.08em
\lightrulewidth=.05em
\cmidrulewidth=.03em
\belowrulesep=.65ex
\belowbottomsep=0pt
\aboverulesep=.4ex
\abovetopsep=0pt
\cmidrulesep=\doublerulesep
\cmidrulekern=.5em
\defaultaddspace=.5em
}

\begin{document}

\title{Phonon dispersion of \mos{}}

\author{Hans Tornatzky}
\email{ht07@physik.tu-berlin.de}
\affiliation{Institut f\"ur Festk\"orperphysik, Technische Universit\"at Berlin Hardenbergstr. 36, 10623 Berlin, Germany}

\author{Roland Gillen}
\affiliation{Department Physik, Friedrich-Alexander-Universit\"at Erlangen-N\"urnberg, Staudtstr. 7, 91058 Erlangen, Germany}

\author{Hiroshi Uchiyama}
\affiliation{Japan Synchrotron Radiation Research Institute (JASRI/SPring-8), 1-1-1 Kouto, Sayo, Hyogo 679-5198 Japan}

\author{Janina Maultzsch}
\email{janina.maultzsch@fau.de}
\affiliation{Department Physik, Friedrich-Alexander-Universit\"at Erlangen-N\"urnberg, Staudtstr. 7, 91058 Erlangen, Germany}

\date{\today}

  \begin{abstract}
  Transition metal dichalcogenides like MoS$_2$, MoSe$_2$, WS$_2$, and WSe$_2$ have attracted enormous interest during recent years. They are van-der-Waals crystals with highly anisotropic properties, which allows exfoliation of individual layers. Their remarkable physical properties make them promising for applications in optoelectronic, spintronic, and valleytronic devices. Phonons are fundamental to many of the underlying physical processes, like carrier and spin relaxation or exciton dynamics. However, experimental data of the complete phonon dispersion relations in these materials is missing. Here we present the phonon dispersion of bulk MoS$_2$ in the high-symmetry directions of the Brillouin zone, determined by inelastic X-ray scattering. Our results underline the two-dimensional nature of MoS$_2$. Supported by first-principles calculations, we determine the phonon displacement patterns, symmetry properties, and scattering intensities. The results will be the basis for future  experimental and theoretical work regarding electron-phonon interactions, intervalley scattering, as well as phonons in related 2D materials.
  
  \end{abstract}

\maketitle

Lattice dynamics constitute one of the most fundamental properties of a crystal, being the basis for mechanical and elastic properties, thermal transport as well as charge-carrier dynamics, phonon-assisted optical excitations and many more. In this view, it is highly desired to have reliable data about the phonon dispersion relation of MoS$_2$, a layered crystal that has boosted the new research field of two-dimensional (2D) materials beyond graphene during recent years~\cite{Mak2010,Xu2014,Novoselov2016,vdw-geim}. This is due to its fascinating physical properties in single-layer form, which it shares with related transition-metal dichalcogenides (TMDCs) like MoSe$_2$, WS$_2$, WSe$_2$, or MoTe$_2$~\cite{poellmann2015,selig2016,phonon-polaritons,lightwave-valleytronics}.
Many of their physical processes  relevant for new applications~\cite{gate-memory,triboelectricity}, such as carrier and exciton dynamics~\cite{ultrafast-carrierdyn,selig2016,ruppert17}, decay of so-called valley polarization~\cite{schmidt2016,miyauchi2018,tornatzky} (the selective population of one of the two inequivalent $K$ points in the Brillouin zone~\cite{Mak2012}), electron-phonon coupling in superconducting states~\cite{kang2018-nv,kang2018}, and relaxation of spins~\cite{spin-relaxation}, crucially depend on phonons.
For example, phonons with considerably large wave vector are required for optical absorption and emission from the indirect band gap in few-layer and bulk TMDCs. They are the relevant source for electron scattering in electron transport~\cite{cui2015} and are expected to play a significant role in the formation of momentum-space indirect interlayer excitons in van-der-Waals heterostructures~\cite{interlayer-excitons,kunstmann-interlayer_excitons}.

In MoS$_2$ and other TMDCs, however, experimental data on the full phonon dispersion are missing. Only one high-symmetry direction of the Brillouin zone in MoS$_2$ has been accessed so far by inelastic neutron scattering (INS)~\cite{Wakabayashi1975}. However, the part most relevant for scattering with large phonon wave vectors $q$ or between the inequivalent points $K$ and $K^\prime$ (i.e., between the valleys) is completely missing and has been addressed by calculations only so far, e.g., in Refs.~\cite{altes-buch,Molina-Sanchez2011,Ataca2012}. The same holds for the other TMDCs, which have  in common, as an obstacle for INS experiments, the in-plane nature of the phonon dispersion and the lack of large single crystals. 

Here we present the entire phonon dispersion relation of MoS$_2$ in the high-symmetry directions $\Gamma -K$, $\Gamma -M$, $K-M$, and $\Gamma -A$ of the Brillouin zone as determined by inelastic X-ray scattering (IXS) experiments. We show the existence of almost degenerate Davydov pairs throughout the Brillouin zone, as well as nearly quadratic dispersion of the out-of-plane acoustic modes (flexural modes), underlining the two-dimensional nature of MoS$_2$. 
Our results are further substantiated by density-functional theory calculations and simulations of the structure factor, which determines the scattering intensities. They are in excellent agreement with the experimental data and reveal the mixing of in-plane and out-of-plane phonon displacement directions inside the Brillouin zone.

\begin{figure}[b]
\includegraphics[trim=10 8 280 22 clip, width=0.48\textwidth]{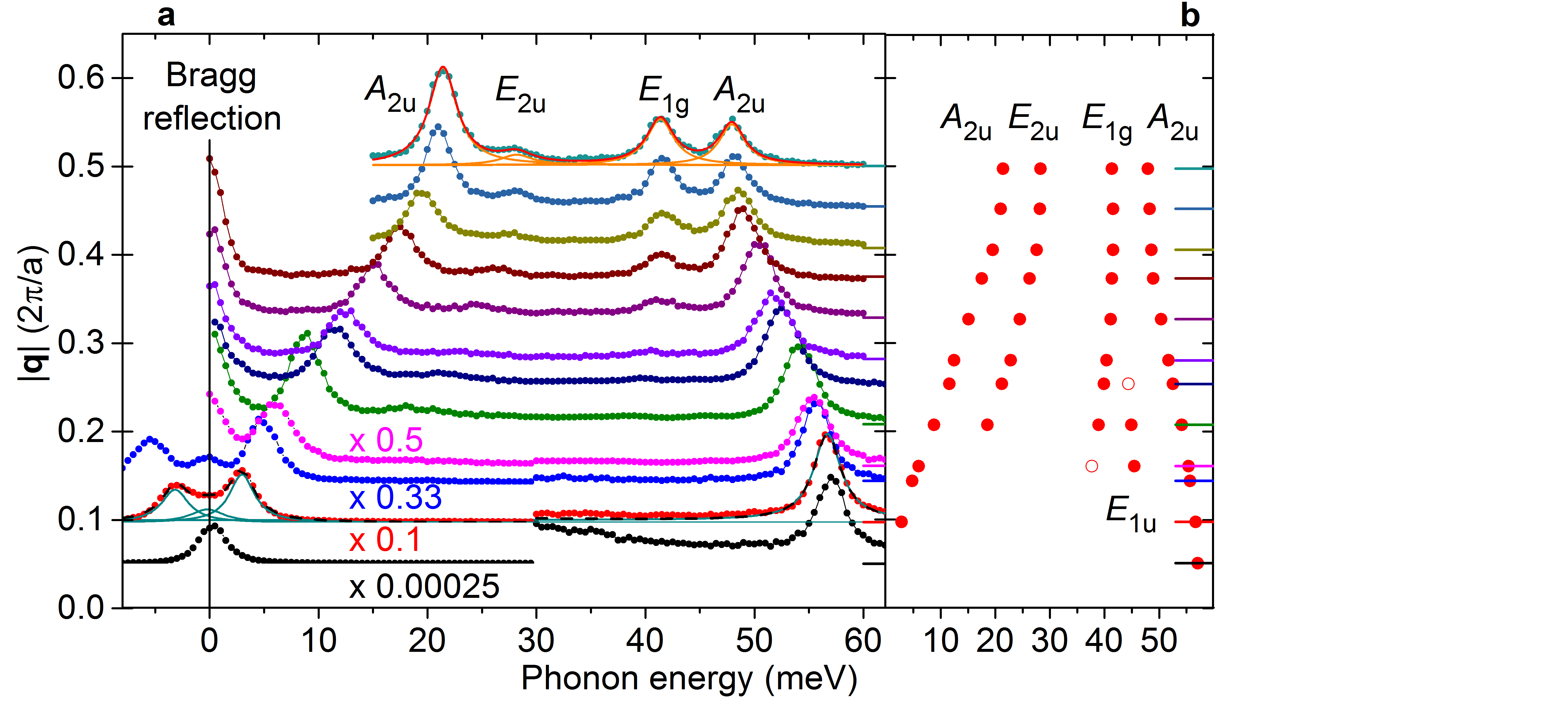}
\caption{ 
 \textbf{a} Experimental IXS spectra of \mos{} along the $\Gamma$-M direction (in the vicinity of the \mbox{(0\,0\,12)} Bragg reflection) with vertical offsets corresponding to the phonon $q$ 
vector. The four spectra closest to $\Gamma$ are scaled by the factor given next to the spectrum. 
 \textbf{b} Extracted peak positions of the spectra shown in \textbf{a}. Peaks are labeled according to their symmetry and  notation at the $\Gamma$ point.
	  }
 \label{fig:spectra}
\end{figure}

2H-\mos{} forms a hexagonal crystal with space group P6$_3$/mmc ($D^4_{6\text{h}}$ in Sch\"onflie\ss{} notation) with six atoms in the unit cell, giving rise to 18 phonon branches. At the $\Gamma$ point, the phonon modes decompose into the irreducible 
representations 
\[
  \Gamma_{2H} = A_{1g} \oplus 2A_{2u} \oplus 2B_{2g} \oplus B_{1u} \oplus E_{1g} \oplus 2E_{1u} \oplus 2E_{2g} \oplus E_{2u}
\]
for the conventional definition of a 120$^\circ$ angle between the in-plane lattice vectors~\cite{jorio-grouptheory,Scheuschner2015}. 
If an angle of $60\,^\circ$ between the in-plane lattice vectors is used, the irreducible representation  $B_{2g}$ is interchanged with  $B_{1g}$  and $B_{1u}$ with $B_{2u}$.
Due to the relatively weak non-covalent coupling of the MoS$_2$ layers, the phonon branches are nearly doubly degenerate at almost all $q$-vectors in 
the Brillouin zone and can be thought of as corresponding to the 9 phonon branches of single-layer \mos{} (three atoms per unit cell, space group 
P$\bar{6}$m2, $D_{3\text{h}}$).

\begin{figure*}
 \includegraphics[width=\textwidth]{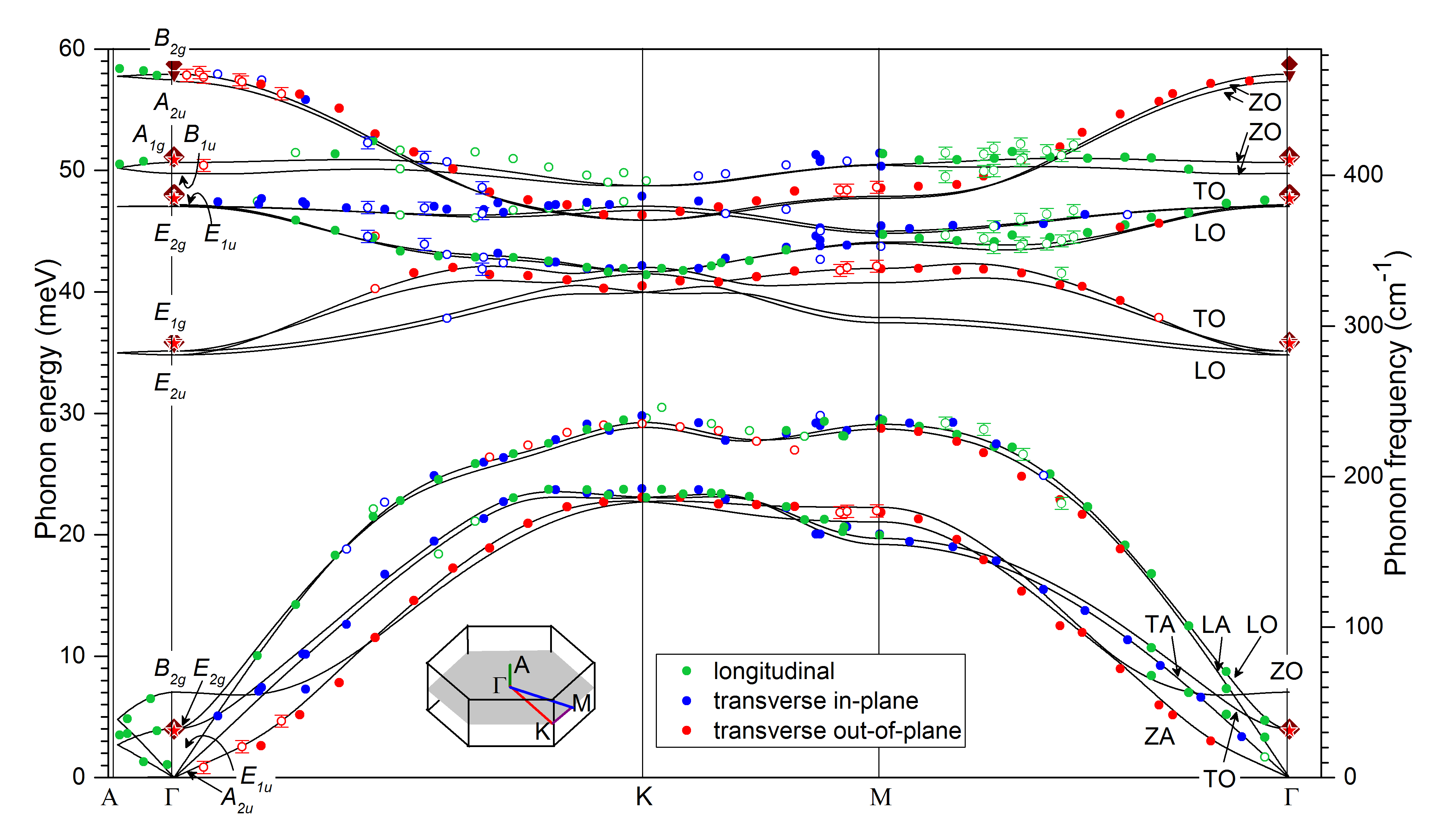}
 \caption{Inelastic X-ray scattering measurements and density-functional perturbation theory calculations of the phonon dispersion of \mos{} along the high 
symmetry directions A-$\Gamma$-K-M-$\Gamma$. Circles in green / blue / red represent measurements probing phonons with an in-plane longitudinal (L) / in-plane transverse (T) / 
out-of-plane transverse (Z) component.  For symbols without an error bar, the error is estimated to be smaller than the symbol size. Open symbols depict peaks with small intensities or larger error. Values at the $\Gamma$ point are 
Raman and IR spectroscopy data from the literature~\cite{Wieting1971, Chen1974, Scheuschner2015} (diamonds) and from Raman measurements on the same sample as used in the IXS experiment (stars).
Phonon branches are labeled by their symmetry within the $D_{6h}$ at the $\Gamma$ point, as well as by their displacement (L, T, Z) and acoustic (A) or optical (O) character. For a symmetry labeling at the $K$ and $M$ points, see Tab.~\ref{tab:phonon_freq}; for a compilation of the eigenvectors see Tables.~S\,I-III in the supplementary material.}
 \label{fig:dispersion}
\end{figure*}

In Fig.~\ref{fig:spectra}\,\textbf{a}, a selection of the measured IXS spectra is plotted. The 
displayed spectra were measured along \mbox{(0\,0+$q$\,12)}, with the absolute value of the phonon wave vector $|{\bf q}|=q=0\ldots0.5$ in units of the reciprocal 
lattice vector. Here, ($h$\,$k$\,$l$) are the Miller indices identifying the scattering vector of the elastically scattered light, \textit{i.e.} a Bragg peak. 
 The extracted peak positions thus represent the dispersion of the out-of-plane transverse modes from $\Gamma$ to $M$, see Fig.~\ref{fig:spectra}\,\textbf{b}.

The IXS measurements allowed us to identify all acoustic and almost all optical phonon branches  along the $A$-$\Gamma$-$K$-$M$-$\Gamma$ directions as shown in Fig.~\ref{fig:dispersion}. IXS data are shown by circles together with  theoretical values from DFT calculations (lines). Phonon energies obtained by Raman spectroscopy of the same sample (stars in Fig.~\ref{fig:dispersion}) 
complement the IXS results with data at the $\Gamma$ point. We find seemless agreement between the $\Gamma$-point frequencies and the IXS data.

\newcommand{\chen}{$^\text{\cite{Chen1974}}$}
\newcommand{\wiet}{$^\text{\cite{Wieting1971}}$}
\newcommand{\chenwiet}{$^\text{\cite{Chen1974,Wieting1971}}$}
\newcommand{\murow}[1]{\multirow{2}{*}{#1}}
\newcommand{\rot}{\footnote{measured with 633\,nm excitation}}
\newcommand{\blau}{\footnote{measured with 457\,nm excitation}}

\begin{table*}[hbt]
 \caption{ Phonon frequencies (given in \cm{}) of \mos{} at the $\Gamma$, $K$ and $M$ high-symmetry points from our DFT calculations and experiments (Raman and IR spectroscopy at $\Gamma$ and IXS at $K$, $M$). The phonon modes are labeled by their irreducible representations in the factor groups $D_{6h}$ ($\Gamma$), $D_{3h}$ ($K$) and $D_{2h}$ ($M$). For a group theory analysis of phonons in MoS$_2$, see also~\cite{jorio-grouptheory}.}
      \label{tab:phonon_freq}
\begin{tabular}{@{}cclrccccccc@{}}
	      \toprule
		\multicolumn{5}{c<{}|}{$\mathbf{\Gamma}$ (space group $D^4_{6h}$)} & \multicolumn{3}{c<{}|}{\textbf{K} ($D^4_{3h}$)} & \multicolumn{3}{c<{}}{\textbf{M} ($D^{17}_{2h}$)}\\
\multicolumn{2}{c<{}}{irr. rep.} &  \multicolumn{1}{l}{\hskip3ex Raman} & \multicolumn{1}{c<{}}{IR} & \multicolumn{1}{c<{}|}{DFT} & 
\multicolumn{1}{c<{}}{irr. rep.} & \multicolumn{1}{c<{}}{\phantom{}IXS} & \multicolumn{1}{c<{}|}{DFT} &
\multicolumn{1}{c<{}}{irr. rep.} & \multicolumn{1}{c<{}}{\phantom{}IXS} & \multicolumn{1}{c<{}}{DFT} \\	
\midrule
  \murow{$E_{1u}$} & TA & \murow{inactive}                        & \murow{$\rightarrow$0}  & \murow{0.0}    & \murow{$E'$}             &  \murow{189}  & \murow{186.0} & $B_{1g}$   & \murow{161} & 154.7 \\
                   & LA &                                         &                         &                &                          &               &               & $B_{2u}$   &             & 158.7 \\
  \murow{$E_{2g}$} & TO & \murow{31 -- 32\rot\chen}               & \murow{inactive}        & \murow{32.2}   & $A_2'$                   &  \murow{239}  & 232.5         & {$A_{g}$}  & \murow{235} & 231.5 \\
                   & LO &                                         &                         &                & $A_1'$                   &               & 235.8         & {$B_{3u}$} &             & 234.6 \\\hline
  $A_{2u}$         & ZA & \murow{inactive}                        & $\rightarrow$0          & 0.0            & \murow{$E''$}            &  \murow{186}  & \murow{183.1} & {$B_{2g}$} & \murow{176} & 179.4 \\
  $B_{2g}$         & ZO &                                         & inactive                & 56.6           &                          &               &               & {$B_{1u}$} &             & 169.7 \\\hline
  \murow{$E_{2u}$} & LO & \murow{inactive}                        &\multirow{4}{*}{inactive}& \murow{280.7}  & \murow{$E''$}            &  \murow{327}  & \murow{322.3} & {$B_{3g}$} &             & 302.1 \\
                   & TO &                                         &                         &                &                          &               &               & {$A_{u}$}  &             & 305.5 \\
  \murow{$E_{1g}$} & LO & \murow{286 -- 289\blau\chenwiet}        &                         & \murow{283.3}  & $A_2''$                  &  334          & 334.5         & {$B_{2g}$} &             & 328.7 \\	
                   & TO &                                         &                         &                & $A_1''$                  &  340          & 339.0         & {$B_{1u}$} & 338         & 338.1 \\\hline
  \murow{$E_{2g}$} & LO & \murow{383 -- 384$^b$\chenwiet}         & \murow{inactive}        & \murow{379.2}  & \murow{$E'$}             &  \murow{334}  & \murow{335.9} & {$A_{g}$}  & \murow{353} & 354.8 \\
                   & TO &                                         &                         &                &                          &               &               & {$B_{3u}$} &             & 355.6 \\
  \murow{$E_{1u}$} & LO & \murow{inactive}                        & \murow{384\wiet}        & 379.4          & $A_2'$                   &  \murow{384}  & 376.7         & {$B_{1g}$} & 361         & 361.4 \\
                   & TO &                                         &                         & 380.4          & $A_1'$                   &               & 379.4         & {$B_{2u}$} & 364         & 362.9 \\\hline	
  $B_{1u}$         & ZO &  inactive                               & \murow{inactive}        & 401.3          & \murow{$E'$}             &  \murow{398}  & \murow{393.0} & {$A_{g}$}  & \murow{410} & 406.6 \\
  $A_{1g}$         & ZO &  408 -- 410$^b$\chenwiet                &                         & 408.5          &                          &               &               & {$B_{3u}$} &             & 407.1 \\\hline	
  $A_{2u}$         & ZO &  \murow{inactive}                       & 470\wiet                & 462.2          & \murow{$E''$}            &  \murow{373}  & \murow{370.1} & {$B_{2g}$} & \murow{392} & 386.0 \\
  $B_{2g}$         & ZO &                                         & inactive                & 467.1          &                          &               &               & {$B_{1u}$} &             & 384.7 \\
   \bottomrule
	\end{tabular}
\end{table*}

We observe nine phonon branches. The calculations show that they are almost doubly degenerate, as expected from the weak interlayer forces. The energy resolution in the IXS experiment, however, does not allow to distinguish these so-called Davydov pairs: Each pair is formed by two phonon modes, where  the two layers forming the bulk unit cell $(i)$ have both the same displacement pattern as the single layer and $(ii)$ the displacement of one of the layers is shifted in phase by $\pi$. At the $\Gamma$ point, one of the bulk modes of such a pair is always even with respect to spatial inversion and the other one is odd~\cite{Wieting1971,Scheuschner2015}. The frequency difference of the two phonons in a Davydov pair is very small if the interaction between the layers is weak. Only the acoustic phonons have fundamentally different behavior: one mode is still acoustic (zero frequency), whereas the other one has finite frequency and corresponds to a rigid-layer vibration at the $\Gamma$ point.

Near the $\Gamma$ point, the phonon modes have well defined displacement direction, i.e., in-plane longitudinal (L), in-plane transverse (T), and out-of-plane transverse (Z). This is seen by the color of the symbols representing the IXS data, which indicates the displacement direction preferentially detected in the given scattering geometry. Towards the $K$ and $M$ points, we observe data points with different colors (i.e., different preferred displacement directions) on the same branch, see for instance the longitudinal acoustic (LA) branch. We interpret this by an increased mixing of the displacement directions for increasing $q$.
This is supported by our calculations of the phonon eigenvectors, see Fig.\,\ref{fig:eigenvectors} for the example of the transverse $E_{1g}$ branch, and Tab.~SI-III in the supplementary material for a  compilation of all 18 eigenvectors at the $\Gamma$, $K$, and $M$ points. Note that the mixing is not limited to the in-plane direction, but includes the out-of-plane modes as well, in contrast to the example of graphite~\cite{Mohr2007}. 
This mixing also explains why some of the phonon branches are only partially observed or show weak signal: for instance, the phonon branch with $E_{1g}$ symmetry at the $\Gamma$ point (in-plane vibration) cannot be observed in the chosen scattering geometry (see also discussion below). However, it gains an out-of-plane component for $q>0$, see Fig.~\ref{fig:eigenvectors}, which results in (weak) IXS signal. 

Furthermore, we observe  a quadratic dispersion of the ZA branch (also called flexural mode) near the $\Gamma$ point, which is typical for two-dimensional atomically thin sheets~\cite{quadratic-ZA} and underlines the 2D nature of the MoS$_2$ layers even within the bulk crystal. As in the case of graphite and graphene~\cite{Maultzsch2004,pisana2007,yan2007,Mohr2007}, the phonon dispersion of bulk MoS$_2$ is thus expected to be indicative of the phonon dispersion in monolayer MoS$_2$ as well as in related TMDCs.

\begin{figure}
\raggedright
 $\Gamma$ ($E_{1g}$) \hspace{5em} $0.5\,\overline{\Gamma M}$ \hspace{9em} $M (B_{1u})$ 

\begin{minipage}[c]{.0528\textwidth}
\rule{3pt}{0pt}
\includegraphics[scale=0.09]{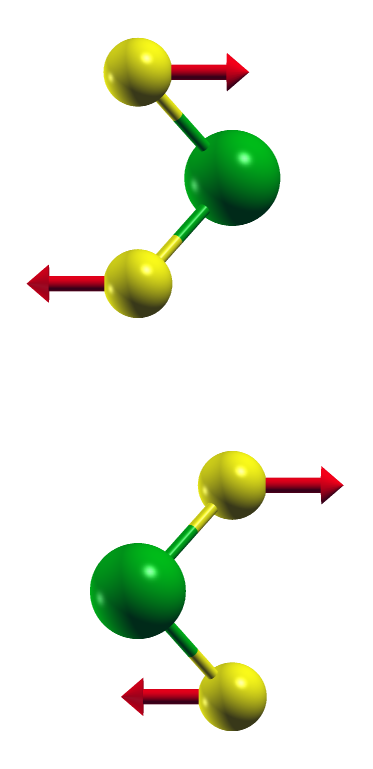}
\end{minipage}
\hspace{5pt}
\begin{minipage}[c]{.264\textwidth}
\includegraphics[scale=0.09]{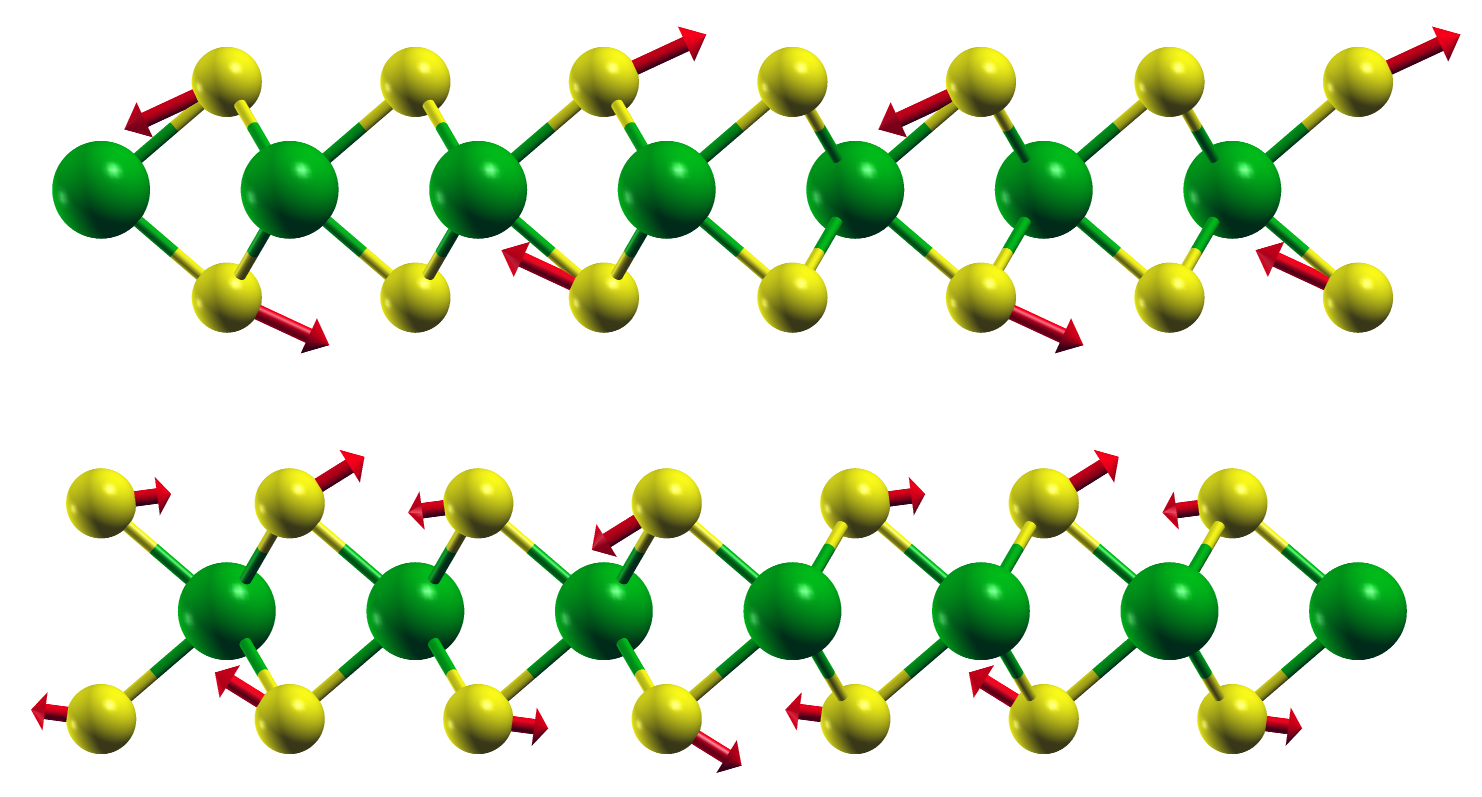}
\end{minipage}
\hspace{5pt}
\begin{minipage}[c]{.12\textwidth}
\includegraphics[scale=0.09]{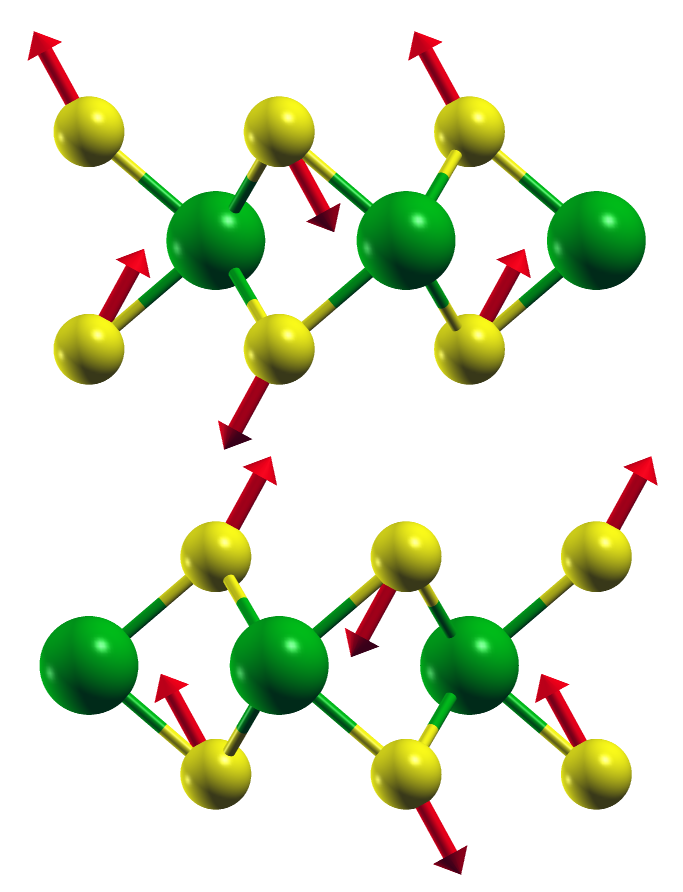}
\end{minipage}
  \caption{Transition of the phonon eigenvectors from the $\Gamma$ to the M point of the branch with {$E_{1g}$ (TO)} symmetry at the $\Gamma$ point. See tables\,S\,I-III in the supplementary material for a complete overview of the phonon eigenvectors at the ${\Gamma}$, ${K}$ and ${M}$ points. 
	  }
  \label{fig:eigenvectors}
\end{figure}

The phonon dispersion obtained from DFT simulations shows overall excellent agreement with the IXS data. As discussed in the following, the challenge for lattice-dynamics simulations in layered materials is the proper description of the effect of non-covalent interactions, which bind the individual layers together. This is particularly relevant for 
the rigid-layer, low-frequency shear and breathing-like modes in the vicinity of the $\Gamma$-point, where the force-constants are small and dominated by contributions from the non-covalent interlayer coupling. 

Therefore, we implemented the contributions from the popular DFT-D3 van-der-Waals corrections~\cite{d3-2} to the dynamical matrix~\cite{d3-dispersion} to the density functional perturbation theory code in the Quantum Espresso package~\cite{qe}. These semi-empirical corrections introduce an additional attractive London-like interatomic potential, which compensates for the underbinding and intrinsic exponential decay of non-classical interactions in the GGA-PBE exchange-correlation approximation we used for our computations. The obtained lattice constants of $a$=3.158\,\AA\space and $c$=12.229\,\AA\space from our PBE+D3 calculations are close to the lattice constants of our MoS$_2$ sample of $a^{\mbox{exp}}$=3.161\,\AA\space and $c^{\mbox{exp}}$=12.297\,\AA, suggesting an excellent description of both covalent and non-covalent interatomic bonding in MoS$_2$. 

The inclusion of non-covalent interactions leads to a noticeable improvement for the values of the low-frequency modes close to the $\Gamma$-point, which are 
significantly underestimated in the GGA-PBE approximation (not shown). Our calculated phonon dispersion is in good quantitative agreement with the IXS 
measurements for the acoustic and low-energy optical modes (Tab.~\ref{tab:phonon_freq}) and correctly describes the small 'bumps' in the dispersion of the Davydov pair of the LA-derived branch [$E_{2g}$ (LO, shear mode) and $E_{1u}$ (LA) at the $\Gamma$ point] along $\Gamma$-$K$-$M$-$\Gamma$, see Fig.~\ref{fig:dispersion}. 
For the higher-frequency modes, our PBE+D3 approach appears to perform slightly less well compared to the Raman measurements and systematically underestimates the 
frequencies of the Raman active modes by a few cm$^{-1}$, see Tab.~\ref{tab:phonon_freq}. 

On the other hand, Molina-S\'anchez~\textit{et al.}~\cite{Molina-Sanchez2011} have shown that phonon dispersions from calculations using the local-density 
approximation (LDA) offer a reasonable qualitative and quantitative description of the available data from the previous INS experiments~\cite{Wakabayashi1975}. 
In general, the predicted LDA frequencies are somewhat higher than those from our PBE+D3 calculations, hence leading to a slightly better agreement between theory and experiment for the high-energy optical modes, but a worse agreement for the acoustic and low-energy optical modes. The good agreement seems consistent for a wide range of layered crystals, but is to a certain extent fortuitous due the intrinsic overbinding of LDA causing a hardening of the predicted phonons at the cost of significantly lower-quality lattice constants. An improved quantitative agreement over the full frequency range of MoS$_2$ and similar layered materials hence requires an exchange-correlation approximation that predicts stronger in-plane covalent bonding than PBE and sufficiently soft interlayer non-covalent interaction.

While we observed the phonon dispersion of eight of the nine Davydov pairs in the experiment, we
were unable to access the almost degenerate branches derived from the two LO modes around 35\,meV ($E_{2u}$ and $E_{1g}$ at the $\Gamma$ point) in the $\Gamma$-$K$ and $\Gamma$-$M$ directions in our scattering geometries. In order to understand this, we simulated the dynamical structure factor~\cite{Baron-strucfact, DHO} using data from our DFT calculations. The simulations suggest that destructive interference of the counter-phase oscillation of the sulfur sublayers in each MoS$_2$ layer cause extinction of the structure factor for all $q$-vectors along the $\Gamma$-$M$ and $\Gamma$-$K$-$M$ directions, if a Bragg peak \mbox{$(h\,k\,0)$} is used.

\begin{figure}
\centering
\textbf{a} ($h$\,$k$\,$l$)=(-2\,4\,2) \hspace{.05\textwidth}  \textbf{b} ($h$\,$k$\,$l$)=(0\,0\,12) \hspace{.02\textwidth} \phantom{.}\\[-10pt]
\includegraphics*[width=0.2021\textwidth]{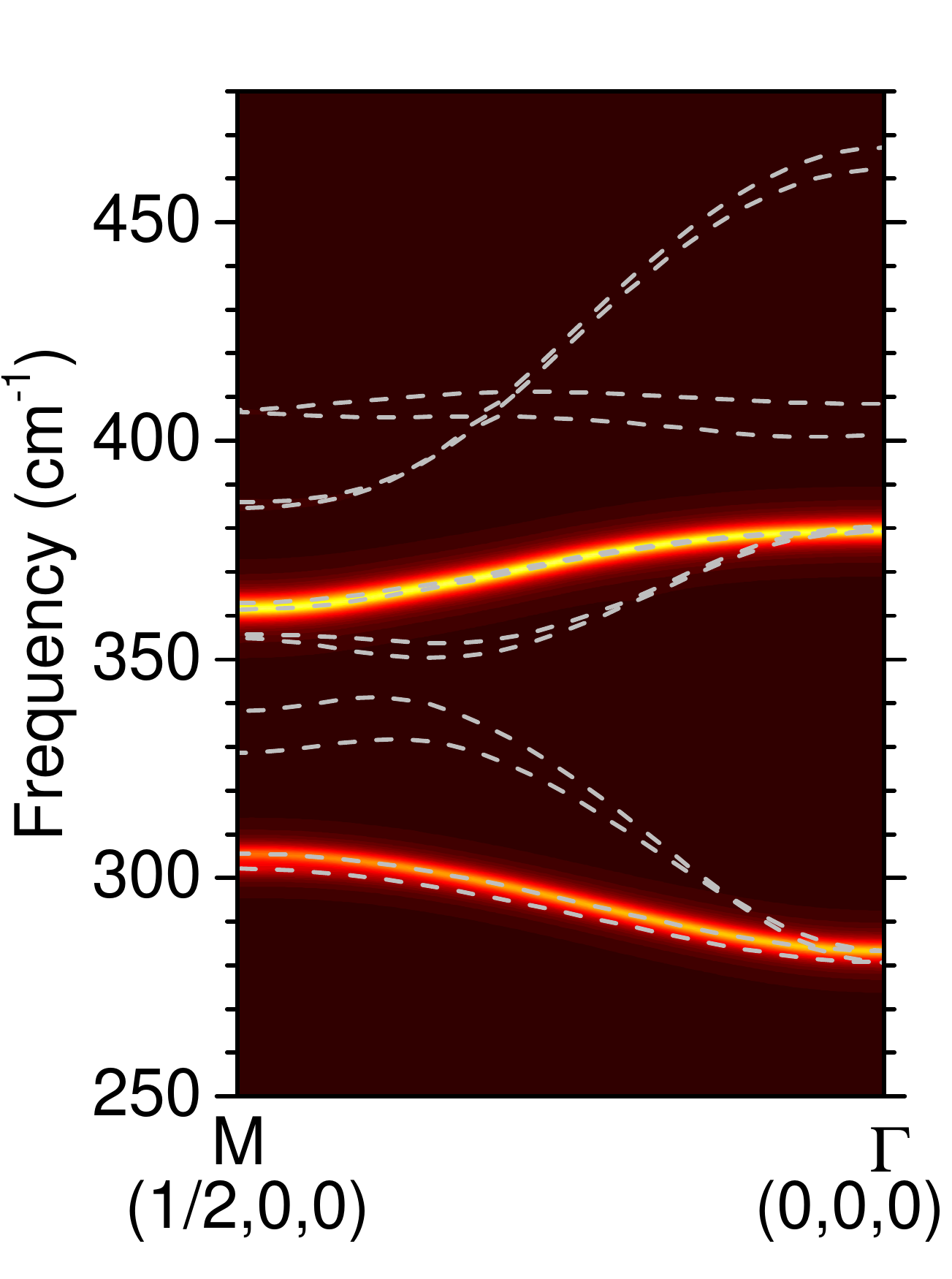}
\includegraphics*[width=0.2721\textwidth]{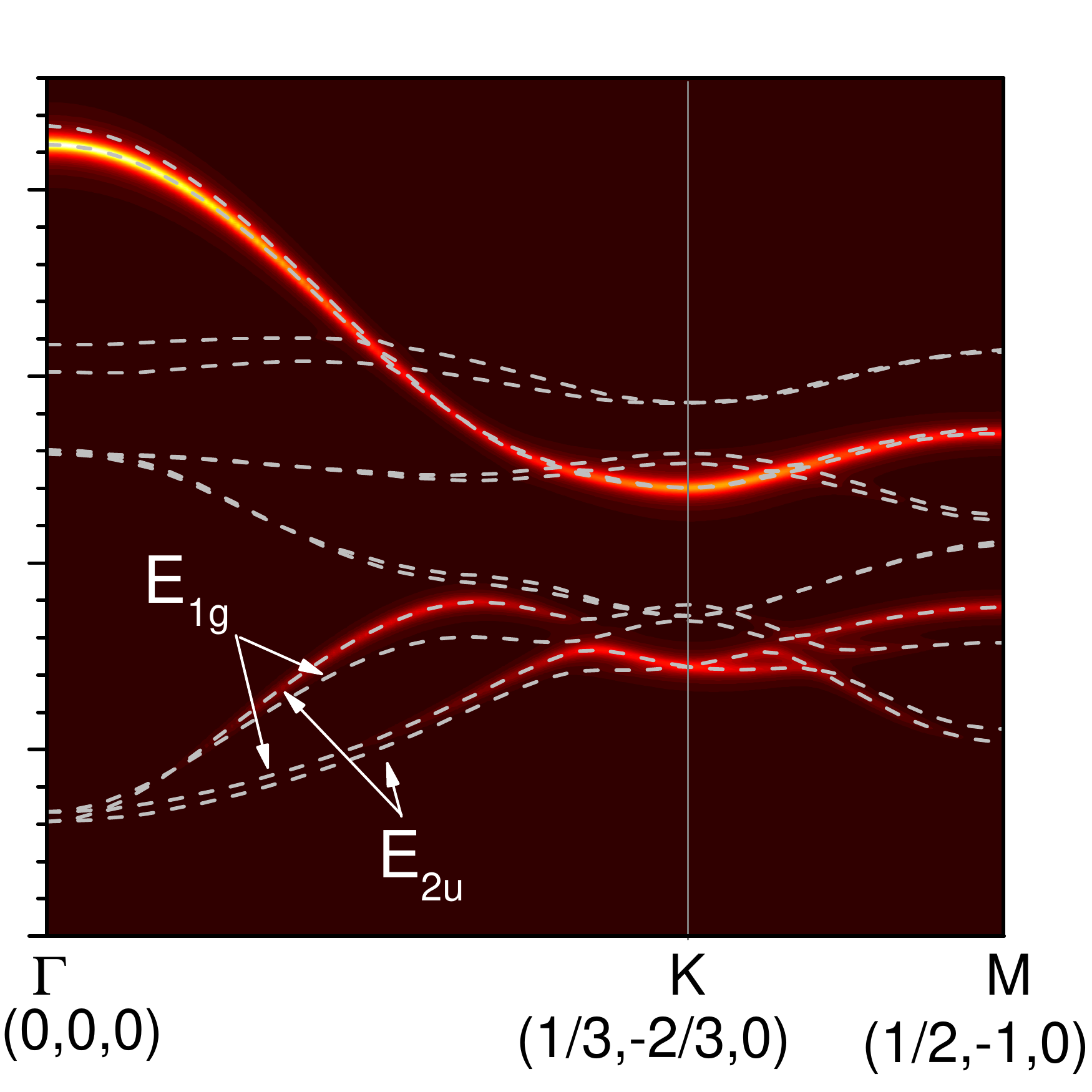}
\caption{\textbf{a} Simulated dynamical structure factor illustrating the possible observation of the branches associated to the longitudinal $E_{1g}/E_{2u}$ modes at the \mbox{(-2\,4\,2)} Bragg peak. \textbf{b} Simulation used for the assignment of the measured out-of-plane phonon energies at the \mbox{(0\,0\,12)} Bragg peak, showing the emergence of significant scattering cross section for the longitudinal $E_{1g}/E_{2u}$ branches around \textit{K}. }
\label{fig:structure_factor_GM_E2u}
\end{figure}

Using a Bragg peak \mbox{$(h\,k\,l)$} with a suitable out-of-plane component $l\neq 0$ should lead to activation of the LO $E_{1g}$ and $E_{2u}$ branches, caused by symmetry breaking of the phase factors from the atomic positions, which lifts the destructive interference. The intensity can be enhanced through a wise choice of the Bragg peak such that it aligns the signs of the contributions from the atomic displacements and of the phase factors from the atomic positions. 
This is illustrated in Fig.~\ref{fig:structure_factor_GM_E2u}\,\textbf{a} for simulated measurements at the Bragg peak \mbox{(-2\,4\,2)} in the direction \mbox{($q$\,0\,0)}. Further, our simulations correctly reproduce the deactivation of the transverse $E_{\text{2u}}$ branch and activation of the longitudinal $E_{1g}$ branch in the vicinity of the \textit{K} point that we observed in our IXS experiments near the (0\,0\,12) peak, see Fig.~\ref{fig:structure_factor_GM_E2u}\,\textbf{b}. This arises from a change of atomic displacement patterns of the longitudinal $E_{2u}$ and $E_{1g}$ modes from a pure in-plane to a pure out-of-plane nature close to the $K$-point, such that these modes behave similarly to the ZO modes. On the other hand, the transverse $E_{2u}$ and $E_{1g}$ modes adopt a mixed out-of-plane/in-plane nature in the middle of the $\Gamma-K$ and $\Gamma-M$ lines (thus coupling to the (0\,0\,12) Bragg peak) but revert back to a pure in-plane character in the vicinity of the $K$ point, see Tab.~SII in the supplementary material.

An overview of the measured phonon frequencies, compared to Raman and IR spectroscopy data at the $\Gamma$ point and to DFT calculation at the $\Gamma$, \textit{K} and \textit{M} points, is given in Tab.~\ref{tab:phonon_freq}. A detailed comparison of our IXS data with measurements performed by electron energy loss spectroscopy (EELS)~\cite{Bertrand1991} and inelastic neutron scattering (INS)~\cite{Wakabayashi1975} and details about our simulations of the dynamical structure factor are given in the supplementary material.

In conclusion, the complete  phonon dispersion relation of MoS$_2$ is determined experimentally in the high-symmetry directions  of the Brillouin zone. In combination with DFT calculations, the data clearly show the 2D character of the lattice vibrations in this layered crystal. Therefore, the results can be immediately transferred to the monolayer form of MoS$_2$ and the vast family of related 2D materials, in particular isostructural TMDCs.  The understanding and engineering of scattering processes involving phonons, such as in electron transport or optical transitions involving the indirect band gap or the two $K$ valleys, will in future be based on the knowledge of the phonon dispersion relation.

\subsection{Acknowledgements}
We thank Daniela Bei\ss e, Marco Haupt, Andreas Ludewig and Wofgang Piper (TU Berlin) for the design and construction of the sample holder and \mbox{Johannes} {Enslin} (TU Berlin) for 
preparative XRD measurements. Computational resources used for the simulations were provided by the HPC of the Regional Computer Centre Erlangen (RRZE). This work was supported by the SPring-8 under proposal number 2017B1738 and by the Deutsche Forschungsgemeinschaft (DFG) within the Cluster of Excellence "Engineering of Advanced Materials" (project EXC 315) (Bridge Funding). \\

\subsection{Methods}

  Inelastic X-ray spectra were recorded at beamline 35XU at the SPring-8 (Japan). The measurements were performed with a photon energy of 17.7935\,keV and a spectral width of $\lesssim3\,$meV (full width of half maximum, FWHM). 
The momentum resolution was set to 0.75\,nm$^{-1}$.
  The focused beam had a spot size of  $75\times 63\,\mu\text{m}^2$, enabling us to select a single crystalline  domain of our bulk \mos{} crystal. A detailed description of the beamline can be found in Ref.~\cite{Baron2000}.
  
  The sample is a synthesized crystal (HQ graphene, Netherlands) with a thickness of about 150\,$\mu$m to match the attenuation length of the 
used X-rays in \mos{}, yielding the best trade-off between high absorption and low scattering in a transmission setup.
All measurements were taken at ambient conditions.

The phonon calculations were performed using density functional perturbation theory (DFPT) module of Quantum Espresso~\cite{qe}, a 12x12x4 k-point sampling and normconserving pseudopotentials~\cite{oncvpsp,pseudodojo} with a cutoff energy of 120\,Ry. Long-range non-covalent interactions in both groundstate and phonon calculations were included through the semi-empirical DFT-D3 correction with Becke-Johnson damping~\cite{d3-2} that we added on top of the PBE exchange-correlation. We used a set of parameters for the D3 corrections that was fitted to successfully reproduce the experimental lattice constants of a wide variety of layered and bulk materials~\cite{exciton-paper, interlayer-excitons,diamondoids}.
A more detailed description of the experimental and theoretical methods can be found in the supplementary material.

\bibliographystyle{naturemag}
\bibliography{bibliography-IXS}

\vspace*{0.5cm}
{\bf Author contributions:}
H.T. and J.M. conceived the experiment, R.G. performed the calculations. All authors performed the IXS experiment, analyzed the data, and discussed the results. H.T., R.G., and J.M. wrote the manuscript with contributions from H.U.

{\bf Competing interests:}
The authors declare no competing interests.

{\bf Materials \& Correspondence:}
Correspondence and material requests should be addressed to H.T.  and J.M.

\end{document}